# Exact projective-steering boundary and sample-size optimization for angularly averaged hyperon–antihyperon states

Weicheng Miao[1], Linzhi Jiang[1,*], and Bingbing Li[2,†]

[1]School of Medical Information and Engineering, Bengbu Medical University, Bengbu 233030, China

[2]Faculty of Engineering, Anhui Sanlian University, Hefei 230601, China

## Abstract

In hyperon–antihyperon production through electron–positron annihilation, quantum steering provides a stricter test of nonclassical spin correlations than entanglement. We determine when steering survives angular averaging and optimize angular selections to reduce the sample size needed to resolve it. In the specified production-frame spin convention, we show that windows symmetric about 90° with even production-angle acceptance yield phase-independent Bell-diagonal states. Applying the necessary and sufficient criterion for steering under arbitrary dichotomic projective measurements determines the exact boundary for this family and reveals a steering region missed by the three-setting Cavalcanti–Jones–Wiseman–Reid (CJWR) criterion. For positive production parameters and uniform full-range acceptance, the steering threshold is $\alpha \simeq 0.36065$. We derive the covariance of weak-decay angular moments, construct fixed linear witnesses, and optimize the angular window. For $J/\psi \to \Lambda\bar{\Lambda}$, steering survives full-range averaging. The statistical model predicts that, after separately optimizing the two angular selections at nominal $z = 5$, the required full-range pair sample decreases from $8.61 \times 10^4$ for CJWR to $2.80 \times 10^4$ for the projective witness, a $67.4\%$ reduction. This advantage persists under the production-acceptance and analyzing-power variations studied.

## 1 Introduction

Quantum steering describes correlations for which the conditional states of one subsystem cannot be reproduced by a local-hidden-state model. Its origins lie in the Einstein–Podolsky–Rosen argument [1] and Schrödinger's analysis of separated systems [2]; its modern formulation places it between entanglement and Bell nonlocality [3–5]. Collider spin measurements provide a setting for probing this hierarchy [6]. Hyperon–antihyperon pairs are particularly useful because parity-violating weak decays analyze the parent spins. Production and decay formalisms [7,8], together with BESIII measurements of polarization and decay parameters [9,10], connect the spin density matrix to observable angular distributions. Building on studies of spin nonlocality, entanglement, and Bell inequalities in hyperon and charmonium decays [11–13], we investigate when the angularly averaged hyperon ensemble retains quantum steering.

Combining events over a finite angular window changes both the reconstructed spin state and the available yield. Kinematic integration is therefore part of collider spin tomography [14,15]. For hyperon pairs, the optimized three-setting Cavalcanti–Jones–Wiseman–Reid (CJWR) inequality [16,17] has been used to determine steering regions at fixed production angle [18]. A narrow angular window can retain strong correlations but select few events; a wider window increases the yield and may dilute the steering signal. We calculate the state averaged with the production rate and acceptance, then evaluate its steering properties. The average uses consistently labeled production-frame components across events, following the frame-dependent ensemble description discussed in Ref. [19]. This construction lets us determine which correlations survive angular averaging and which angular window is most economical for detecting them.

An exact boundary resolves the steerability of states beyond the sufficient tests supplied by finite-setting inequalities [20]. We show that windows symmetric about transverse production, with even production-angle acceptance, cancel the local polarizations and phase-dependent off-diagonal correlations. The average is a Bell-diagonal state. The geometry of steering ellipsoids and steering outcomes [21,22] provides tools for analyzing this family. The local-hidden-state construction of Jevtic et al. [23], proved optimal by Nguyen and Vu [24], yields a necessary and sufficient condition for steering under arbitrary dichotomic projective measurements. The critical-radius framework [25] and its equivalent sphere-integral representation, applied to top-quark pairs in Ref. [26], give practical forms of this condition. We apply it to obtain the exact projective-steering boundary of the averaged hyperon family and identify the steering region missed by CJWR.

The statistical significance of a witness depends on its estimator variance as well as its violation [27]. In weak-decay reconstruction, the analyzing powers control the statistical resolution, while angular selection fixes the fraction of retained pairs. We derive the complete covariance of the daughter-angle moment estimator and construct fixed linear tangent witnesses for the exact projective functional and for CJWR. These witnesses have explicit means and variances and avoid the state-fit bias associated with positivity constraints [28]. Combining their variances with the accepted yield gives the number of full-range reconstructed pairs required for a chosen statistical target.

Contact author: 2021064@bbmu.edu.cn

Contact author: li_b_b@163.com

We minimize this count over the angular window, using the same estimator and counting convention for both criteria.

For uniform full-range acceptance, we obtain the positive-parameter steering threshold $\alpha \simeq 0.36065$, below the three-setting threshold. For $J/\psi \to \Lambda\overline{\Lambda}$, steering survives full angular averaging in a region missed by CJWR. The statistical model predicts that separately optimizing the two fixed-witness designs at nominal $z = 5$ reduces the full-range pair requirement from $8.61 \times 10^4$ for CJWR to $2.80 \times 10^4$ for the projective witness, a 67.4% reduction. The projective witness also favors a wider angular window. Variations of the acceptance shape and analyzing powers show how this gain depends on the experimental inputs.

## 2 Production state and weak-decay observables

We describe unpolarized $e^+e^- \to \psi \to Y\overline{Y}$ with two parity-conserving production amplitudes in the massless-electron approximation. Define

$$\begin{aligned} x = \cos\theta, \quad s &= \sqrt{1-x^2}, \\ d &= 1+\alpha x^2, \\ \beta &= \sqrt{1-\alpha^2}\sin\Delta\Phi, \\ \gamma &= \sqrt{1-\alpha^2}\cos\Delta\Phi. \end{aligned} \tag{1}$$

The scattering angle $\theta$ is measured between the electron beam and the outgoing hyperon. The production parameter $\alpha$ satisfies $-1 \le \alpha \le 1$, and $\Delta\Phi$ is the relative amplitude phase. Detector acceptance enters the event weights separately. We normalize the state on the nonzero-rate domain, excluding configurations with $d = 0$.

For each event in the pair center-of-mass frame, choose $\hat{\mathbf{z}} = \hat{\mathbf{p}}_Y$, $\hat{\mathbf{y}} = (\hat{\mathbf{p}}_Y \times \hat{\mathbf{p}}_e)/|\hat{\mathbf{p}}_Y \times \hat{\mathbf{p}}_e|$, and $\hat{\mathbf{x}} = \hat{\mathbf{y}} \times \hat{\mathbf{z}}$. Daughter four-momenta are transformed to the respective parent rest frames by rotation-free boosts along the parent momenta. Their spatial directions are resolved using the same oriented axis labels on both sides. At $x = \pm 1$, the production plane is undefined; formulas involving its transverse axes are understood by continuity.

In this convention [7,8], the normalized state is

$$\rho = \frac{1}{4}\left[I \otimes I + (\mathbf{p}\cdot\boldsymbol{\sigma}) \otimes I + I \otimes (\mathbf{p}\cdot\boldsymbol{\sigma}) + \textstyle\sum_{i,j} C_{ij}\, \sigma_i \otimes \sigma_j\right], \tag{2}$$

with

$$\mathbf{p} = \left(0, \frac{\beta x s}{d}, 0\right), \qquad C = \frac{1}{d}\begin{pmatrix} 1-x^2 & 0 & \gamma x s \\ 0 & -\alpha(1-x^2) & 0 \\ \gamma x s & 0 & \alpha + x^2 \end{pmatrix}. \tag{3}$$

The only possibly nonzero eigenvalues are $(1+q)/2$ and $(1-q)/2$, where $q = \alpha(1-x^2)/d$. The bound $|q| \le 1$ establishes positivity throughout the physical parameter range, including negative $\alpha$. Appendix A gives the pointwise X-state representation.

Let $a_Y$ and $a_{\overline{Y}}$ denote the signed analyzing powers of the selected decay channels and let $\eta = a_Y a_{\overline{Y}}$. For daughter unit directions $\mathbf{n}$ and $\mathbf{m}$, the normalized joint decay-angle density is

$$w(\mathbf{n}, \mathbf{m} \mid x) = \frac{1}{(4\pi)^2}\left[1 + a_Y \mathbf{p}\cdot\mathbf{n} + a_{\overline{Y}} \mathbf{p}\cdot\mathbf{m} + \eta\, \mathbf{n}^{\mathrm{T}} C \mathbf{m}\right]. \tag{4}$$

Equation (4) is normalized per pair of solid angles and connects the production tensor to measurable decay directions. Production-angle acceptance enters the state average in Sect. 4, and the detector response enters the decay likelihood discussed in Sect. 8.

## 3 Pointwise steering witnesses

For qubit projective observables $A_k = \mathbf{a}_k \cdot \boldsymbol{\sigma}$ and $B_k = \mathbf{b}_k \cdot \boldsymbol{\sigma}$, the three-setting CJWR inequality is [16]

$$\begin{aligned} f_3 &= \frac{1}{\sqrt{3}}\left|\textstyle\sum_{k=1}^{3} \mathbf{a}_k^{\mathrm{T}}\, C \mathbf{b}_k\right| \le 1, \\ \mathbf{b}_k \cdot \mathbf{b}_l &= \delta_{kl}. \end{aligned} \tag{5}$$

Optimizing Alice's unit vectors freely and Bob's orthonormal triad gives [17]

$$F_3(\rho) = \| C \|_{\mathrm{F}} = \sqrt{\mathrm{Tr}(C^{\mathrm{T}} C)}. \tag{6}$$

Appendix A gives a constructive proof. For fixed production axes, the witness becomes

$$G_3 = \frac{C_{xx} - C_{yy} + C_{zz}}{\sqrt{3}} = \frac{1+2\alpha-\alpha x^2}{\sqrt{3}(1+\alpha x^2)}. \tag{7}$$

For $\alpha \ge 0$, this equals the sum of the principal singular correlations divided by $\sqrt{3}$, and $G_3 \le F_3$. It is phase independent because its three chosen diagonal correlations do not contain the interference terms.

For the full tensor,

$$F_3^2 = 1 + \frac{2\alpha^2(1-x^2)^2 - 2\beta^2 x^2 (1-x^2)}{(1+\alpha x^2)^2}. \tag{8}$$

Equation (8) reproduces Ref. [18]. For $\alpha > 0$ and $|x| < 1$, its phase-resolved violation region is

$$x^2 < \frac{\alpha^2}{\alpha^2 + (1-\alpha^2)\sin^2\Delta\Phi}. \tag{9}$$

The common violation region over all phases is $|x| < \alpha$, with the maximizing settings evaluated separately at each phase. The fixed-axis prescription satisfies $G_3 > 1$ when

$$\alpha > \frac{\sqrt{3}-1}{2}, \qquad x^2 < \frac{1+2\alpha-\sqrt{3}}{\alpha(1+\sqrt{3})}. \tag{10}$$

Figure 1 displays the witnesses and their angular boundaries. At transverse production, $F_3 = \sqrt{1+2\alpha^2}$ and $G_3 = (1+2\alpha)/\sqrt{3}$. These fixed-angle results set the reference for the angularly averaged states studied below.

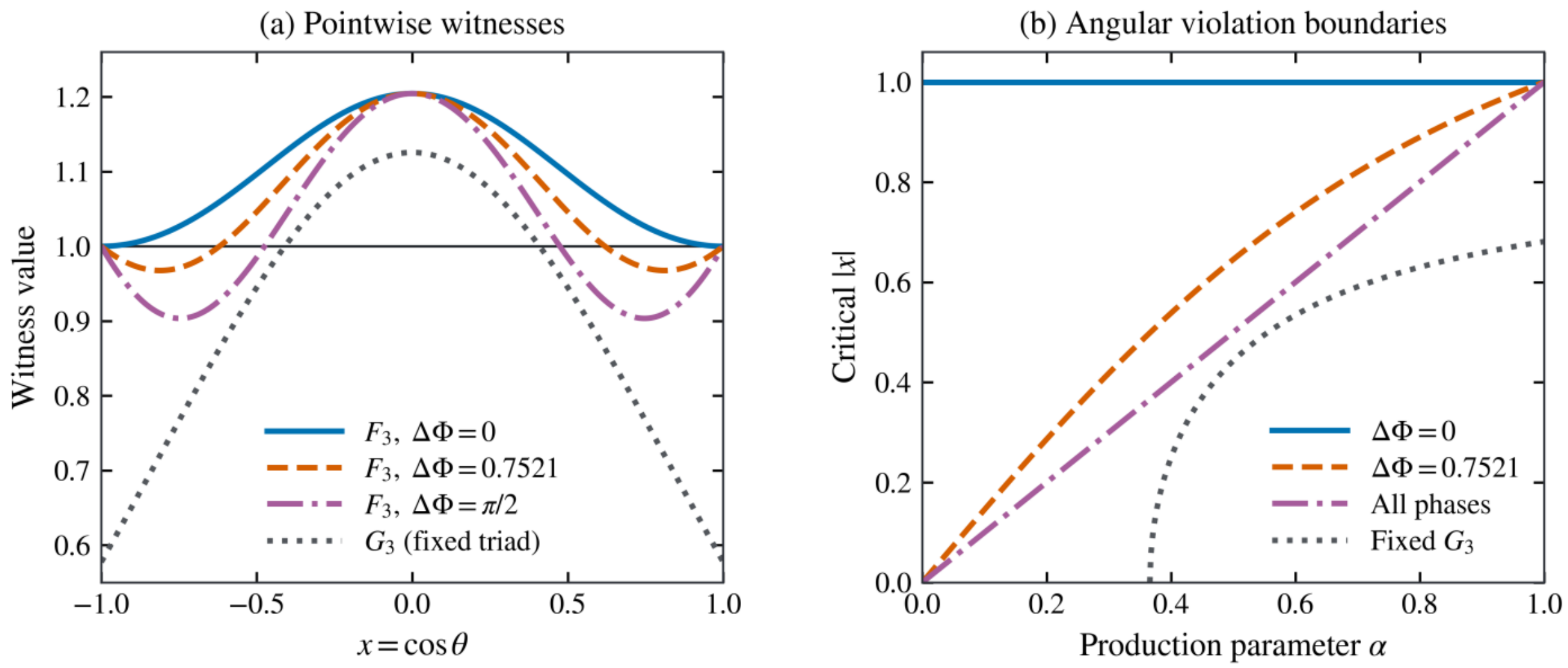

Fig. 1 Pointwise witnesses and their angular boundaries. Panel (a) uses $\alpha = 0.4748$ and the three indicated phases; the dotted curve is the fixed-triad value. Panel (b) gives the critical $|x|$ for the phase-resolved optimum, its all-phase guarantee, and the fixed prescription. The fixed-prescription curve is absent where no violation is possible. These pointwise results serve as the benchmark for the finite-bin analysis.

## 4 Angular averaging and acceptance moments

### 4.1 Definition of the averaged ensemble

Let $B$ be an interval in $x$ and let $\epsilon(x) \geq 0$ be its production-angle acceptance. Identify the labeled components of Eq. (3) across events and define

$$\overline{\rho}_B = \frac{\int_B dx\, \epsilon(x) d(x) \rho(x)}{\int_B dx\, \epsilon(x) d(x)}. \tag{11}$$

Equation (11) averages the components of $C(x)$ defined by the eventwise production axes [19]. For another choice of axes, the corresponding kinematics-dependent local rotations enter inside the integral. This applies to both a laboratory-fixed basis and the pointwise X basis of Appendix A. The prescription links the averaged state to daughter directions resolved along the production axes.

The nonlinear functional is evaluated after the linear tensor average:

$$F_3(\overline{\rho}_B) = \| \overline{C}_B \|_{\mathrm{F}} \leq \frac{\int_B dx\, \epsilon d\, \|C(x)\|_{\mathrm{F}}}{\int_B dx\, \epsilon d}. \tag{12}$$

The Frobenius norm is invariant under local rotations. Equation (12) therefore implies that a violation by the averaged state requires a violation by at least one fixed-angle state in the selected bin.

### 4.2 Reduction to two acceptance moments

For $I_0 = \int_B \epsilon(x)\, dx > 0$, define

$$u = \frac{1}{I_0} \int_B dx\, \epsilon(x) x^2, \qquad v = \frac{1}{I_0} \int_B dx\, \epsilon(x) x \sqrt{1 - x^2}. \tag{13}$$

The moment $u$ is the acceptance-weighted mean of $x^2$, whereas ordinary event averages also contain the production factor $d(x)$. Cauchy–Schwarz gives $0 \leq u \leq 1$ and $v^2 \leq u(1-u)$. For support restricted to $[-L, L]$, one also has $u \leq L^2$.

Multiplying Eq. (3) by the production factor before integration gives

$$\overline{C} = \frac{1}{1+\alpha u} \begin{pmatrix} 1-u & 0 & \gamma v \\ 0 & -\alpha(1-u) & 0 \\ \gamma v & 0 & \alpha + u \end{pmatrix}, \qquad \overline{\mathbf{p}} = \left(0, \frac{\beta v}{1+\alpha u}, 0\right). \tag{14}$$

Thus

$$F_3(\overline{\rho})^2 - 1 = \frac{2\left[(1-u)(\alpha^2 - u) + \gamma^2 v^2\right]}{(1+\alpha u)^2}. \tag{15}$$

Appendix B gives the algebra. An even acceptance in a symmetric interval has $v = 0$, eliminating the local polarizations and all phase dependence of the averaged tensor. In this case

$$F_3(\overline{\rho}) > 1 \quad \Leftrightarrow \quad u < \alpha^2, \qquad 0 \leq u < 1. \tag{16}$$

The asymmetry term $\gamma^2 v^2$ increases the witness at fixed $u$. Changing the acceptance shifts both moments; Eq. (15) gives their combined effect.

For uniform acceptance on $[-L, L]$, with $0 < L \leq 1$, one has $u = L^2/3$ and

$$F_3(\overline{\rho}_L)^2 = \frac{(1+\alpha^2)(1 - L^2/3)^2 + (\alpha + L^2/3)^2}{(1 + \alpha L^2/3)^2}. \tag{17}$$

On the positive branch, CJWR violation requires $L < \sqrt{3}\,\alpha$ within the domain $L \leq 1$. The limit $L \to 0$ approaches the transverse state, with $L = 0$ marking the zero-yield endpoint.

## 5 Exact projective steering of the averaged state

### 5.1 Bell-diagonal family and exact criterion

When $v = 0$, write the diagonal tensor as $T = \mathrm{diag}(c_1, c_2, c_3)$, where

$$\begin{aligned} c_1 &= \frac{1-u}{1+\alpha u}, \\ c_2 &= -\frac{\alpha(1-u)}{1+\alpha u}, \\ c_3 &= \frac{\alpha+u}{1+\alpha u}. \end{aligned} \tag{18}$$

The state has maximally mixed marginals and is Bell diagonal. Its eigenvalues are

$$\{\lambda_i\} = \frac{\{1+\alpha,\, u(1+\alpha),\, (1-\alpha)(1-u),\, 0\}}{2(1+\alpha u)}. \tag{19}$$

The steering-ellipsoid construction [23] and its optimality proof [24] supply the exact criterion for this Bell-diagonal family. In the critical-radius formulation [25], a nonsingular $T$ has

$$\begin{aligned} R_{\mathrm{crit}}(T) &= 2\pi N_T |\mathrm{det} T|, \\ N_T^{-1} &= \int_{S^2} \frac{d\Omega}{(\mathbf{n}^{\mathrm{T}} T^{-2} \mathbf{n})^2}. \end{aligned} \tag{20}$$

The state is projectively steerable if and only if $R_{\mathrm{crit}} < 1$. The equivalent sphere-integral form is [26]

$$\begin{aligned} W_{\mathrm{P}}(T) \equiv \frac{1}{R_{\mathrm{crit}}(T)} &= \frac{1}{2\pi} \int_{S^2} d\,\Omega \ \| T\mathbf{n} \|, \\ W_{\mathrm{P}} &> 1. \end{aligned} \tag{21}$$

For vanishing local polarizations, Eq. (21) is necessary and sufficient for projective steering. For a general correlation tensor $C$, it provides a sufficient witness. Appendix C derives the local-hidden-state bound, the relation between the two integrals, and a one-dimensional elliptic-integral expression for numerical evaluation.

### 5.2 Hierarchy of averaged spin correlations

For comparison with Bell nonlocality [29], we use the Clauser–Horne–Shimony–Holt (CHSH) inequality [30]. For $0 < \alpha \leq 1$ and $0 \leq u < 1$, $|c_2|$ is no greater than either $c_1$ or $c_3$. The Horodecki CHSH criterion [31] therefore gives

$$\begin{aligned} M_{\mathrm{CHSH}} - 1 &= c_1^2 + c_3^2 - 1 \\ &= \frac{(1-u)[\alpha^2(1+u) - 2u]}{(1+\alpha u)^2}, \end{aligned} \tag{22}$$

with $B_{\mathrm{CHSH}} = 2\sqrt{M_{\mathrm{CHSH}}}$. CHSH violation requires $u < \alpha^2/(2-\alpha^2)$. The largest Bell eigenvalue in Eq. (19) gives the concurrence [32]

$$\begin{aligned} \mathrm{C}(\overline{\rho}) &= \frac{\alpha(1-u)}{1+\alpha u}, \\ \alpha > 0, &\quad 0 \leq u < 1. \end{aligned} \tag{23}$$

The additional projective-steering boundary is the interior solution

$$W_{\mathrm{P}}[T(\alpha, u_{\mathrm{P}})] = 1. \tag{24}$$

For $0 < \alpha < 1$ this solution is unique. To see this, set $r = (1-u)/(1+\alpha u)$, so $T = \mathrm{diag}[r, -\alpha r, 1-(1-\alpha)r]$. The integral in Eq. (21) is strictly convex along this line. At $r = 0$, it equals unity with derivative $-(1-\alpha) < 0$; at $r = 1$, it exceeds unity for every $\alpha > 0$. Hence it crosses unity once in $0 < r < 1$, in addition to the separable endpoint.

For any tensor, its singular values also imply

$$F_3(C) \leq W_{\mathrm{P}}(C) \leq \frac{2}{\sqrt{3}} F_3(C). \tag{25}$$

The lower inequality is strict when at least two singular values are nonzero. Applying it at $u = \alpha^2$, where $F_3 = 1$, establishes the ordering

$$\begin{aligned} \frac{\alpha^2}{2-\alpha^2} = u_{\mathrm{CHSH}} < u_{\mathrm{CJWR}} = \alpha^2 < u_{\mathrm{P}} < u_{\mathrm{ent}} = 1, \\ 0 < \alpha < 1. \end{aligned} \tag{26}$$

Figure 2 displays the regions ordered by Eq. (26). All states with $u < 1$ remain entangled on the positive production branch. The projective boundary separates steerable from unsteerable states, while the CHSH curve classifies violation of that Bell inequality. For Bell-diagonal states, steering with two projective settings is equivalent to CHSH violation [33]; Eq. (21) allows arbitrary projective settings.

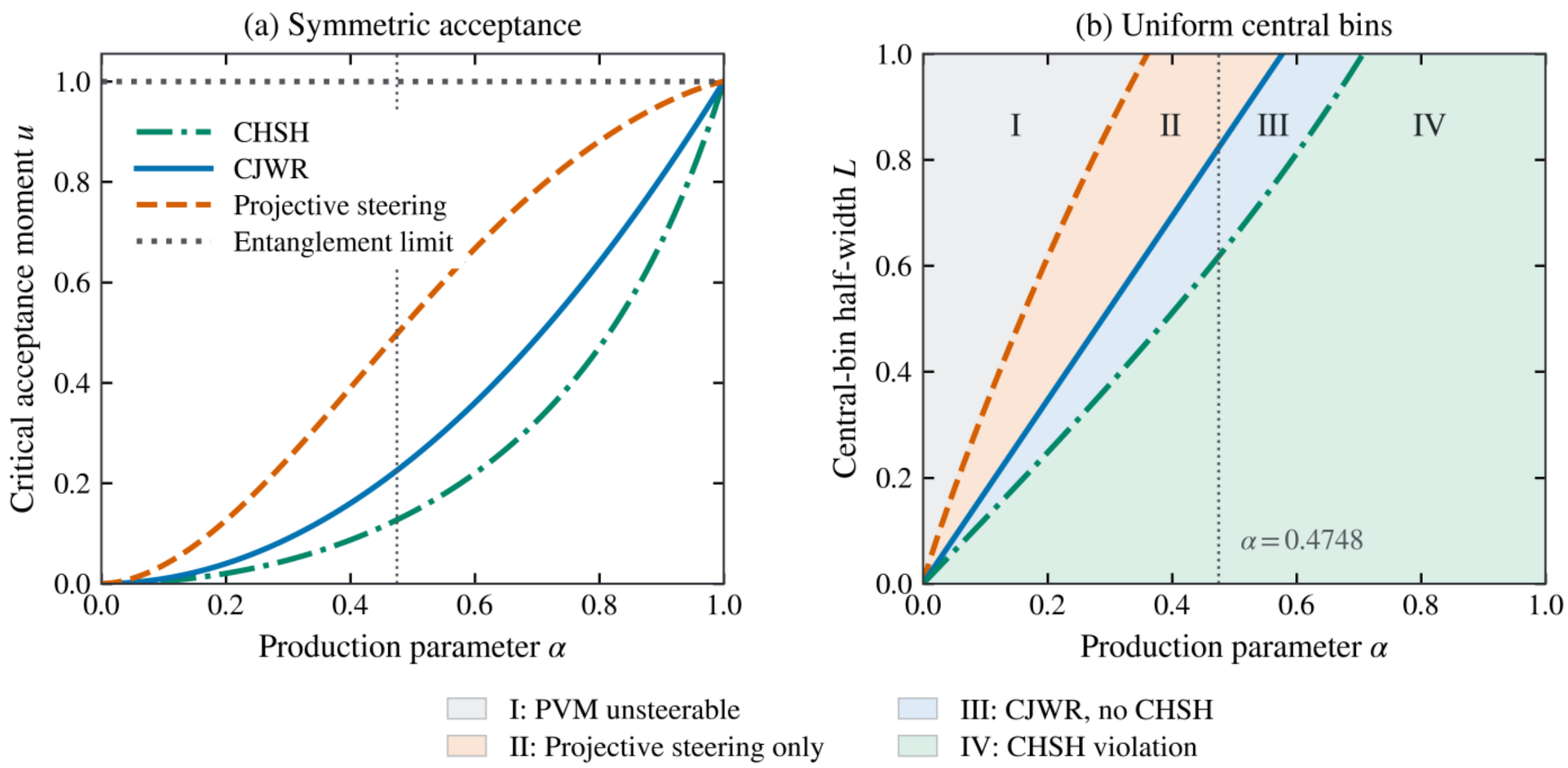

Fig. 2 Hierarchy of symmetrically averaged states on the positive production branch. Panel (a) compares the CHSH, CJWR, exact projective-steering, and entanglement boundaries in the acceptance moment $u$. Panel (b) shows uniform central bins, $u = L^2/3 \leq 1/3$. Regions I–IV distinguish projective unsteerability, projective steering without CJWR violation, CJWR violation without CHSH violation, and CHSH violation. PVM denotes a projective measurement. The endpoint $L = 0$ is the zero-yield limit. The vertical dotted line marks $\alpha = 0.4748$; at $\alpha = 0$ , the averaged states are separable.

For uniform central bins, the CHSH and CJWR critical half-widths are

$$L_{\mathrm{CHSH}} = \frac{\sqrt{3}\,\alpha}{\sqrt{2-\alpha^2}}, \qquad L_{\mathrm{CJWR}} = \sqrt{3}\,\alpha. \tag{27}$$

Substituting $u = L^2/3$ into Eq. (24) determines the production-parameter boundary $\alpha_{\mathrm{P}}(L)$. For the full angular range, $L = 1$,

$$\alpha_{\mathrm{P}} = 0.36064598\ldots, \qquad \alpha_{\mathrm{CJWR}} = \frac{1}{\sqrt{3}}, \qquad \alpha_{\mathrm{CHSH}} = \frac{1}{\sqrt{2}}. \tag{28}$$

The full-range ensemble is therefore projectively steerable throughout $0.36065 < \alpha \leq 1/\sqrt{3}$ , extending beyond the region detected by the three-setting criterion.

## 6 Statistical sensitivity from daughter-angle moments

### 6.1 Complete covariance of the reconstructed tensor

We reconstruct the spin correlation tensor from weighted decay-angle moments [34–36]. For a predetermined bin containing $N$ independent pairs, uniform daughter-angle acceptance, and known $\eta \neq 0$, define

$$\hat{C}_{ij} = \frac{1}{N}\sum_{r=1}^{N} Y_{ij}^{(r)}, \qquad Y_{ij}^{(r)} = \frac{9}{\eta} n_i^{(r)} m_j^{(r)}. \tag{29}$$

Equation (4) gives $\mathrm{E}[n_i m_j] = \eta \bar{C}_{ij}/9$. Every polarization or correlation correction to the fourth daughter-angle moment contains an odd spherical moment and vanishes. Consequently,

$$\mathrm{E}[Y_{ij} Y_{kl}] = \frac{9}{\eta^2}\delta_{ik}\delta_{jl}, \qquad \mathrm{Cov}(\hat{\mathbf{c}}) = \frac{1}{N}\left(\frac{9}{\eta^2} I_9 - \mathbf{c}\mathbf{c}^{\mathrm{T}}\right), \tag{30}$$

where $\mathbf{c} = \mathrm{vec}\bar{C}$. Equation (30) gives the covariance of all nine tensor entries. The outer product describes correlated estimation errors, including those present when the mean tensor is diagonal.

### 6.2 Functionals and pre-specified linear witnesses

Write $J(C) = F_3(C)$ or $W_{\mathrm{P}}(C)$ and let $K_J(C) = \partial J/\partial C$. Their gradients are

$$K_F(C) = \frac{C}{F_3(C)}, \qquad K_{\mathrm{P}}(C) = \frac{1}{2\pi}\int_{S^2} d\Omega\, \frac{(C\mathbf{n})\mathbf{n}^{\mathrm{T}}}{\|C\mathbf{n}\|}. \tag{31}$$

Both functionals are convex and homogeneous of degree one, so $K_J(C){:}C = J(C)$, with $A{:}B = \mathrm{Tr}(A^{\mathrm{T}}B)$. The delta method gives

$$\mathrm{Var}[J(\hat{C})] = \frac{1}{N}\left[\frac{9}{\eta^2}\,\|\, K_J(C)\,\|_{\mathrm{F}}^2 - J(C)^2\right] + o(N^{-1}). \tag{32}$$

For CJWR, this reduces to $(9/\eta^2 - F_3^2)/N$. At a diagonal $T$, $K_{\mathrm{P}}$ is diagonal by parity, and its three entries follow from a single sphere quadrature.

To implement the same leading sensitivity with a linear statistic, fix a design tensor $C_0$ from the production model or an independent design sample and form

$$\hat{S}_J = K_J(C_0):\hat{C}, \quad K_J(C_0):C \leq J(C). \tag{33}$$

Since either $J(C)$ is at most unity for an unsteerable state, $\hat{S}_J$ estimates a linear witness with bound one. At the design state its mean is $J(C_0)$. With the decay density in Eq. (4) and uniform daughter-angle acceptance, its exact finite-$N$ variance is

$$\mathrm{Var}(\hat{S}_J) = \frac{1}{N}\left[\frac{9}{\eta^2}\parallel K_J(C_0)\parallel_\mathrm{F}^2 - (K_J(C_0):C)^2\right]. \tag{34}$$

The sample mean estimates the fixed linear witness without bias. Appendix C gives its event weights and the leading bias of the nonlinear estimate $J(\hat{C})$. The numerical results use the benchmark tensor to set the witness weights before evaluating the sample.

### 6.3 Selected yield and external analyzing-power uncertainty

At the design state, define

$$A_J = \frac{9}{\eta^2}\parallel K_J\parallel_\mathrm{F}^2 - J^2, \quad N_{\mathrm{sel},J}(z) = \frac{z^2 A_J}{(J-1)^2}, \qquad J > 1. \tag{35}$$

Here $z$ measures the expected distance above the witness bound in units of the model standard deviation. This signal-to-noise target sets the planning sample size. A discovery significance is calibrated from the null distribution of the fixed statistic.

For uniform production-angle efficiency, the fraction entering the central bin is

$$f_\mathrm{acc}(L) = \frac{L(1+\alpha L^2/3)}{1+\alpha/3}, \quad N_{\mathrm{full},J}(L) = \frac{N_{\mathrm{sel},J}(L)}{f_\mathrm{acc}(L)}. \tag{36}$$

We minimize $N_\mathrm{full}$ to balance the stronger correlations in a narrow window against the larger yield in a wide window. The count rises both near zero width, where the yield vanishes, and near a witness boundary, where the violation vanishes. It refers to reconstructed decay-channel pairs over the full angular range before the central selection. The corresponding produced-resonance yield is obtained by dividing by the production branching fraction, both decay branching fractions, and the reconstruction efficiency.

An externally determined analyzing-power product with relative uncertainty $\delta_\eta = \sigma_\eta/|\eta|$, independent of the tested sample, adds $J^2\delta_\eta^2$ to the leading variance. The corresponding planning expression becomes

$$N_{\mathrm{full},J}(L,\delta_\eta) = \frac{z^2 A_J}{f_\mathrm{acc}(L)[(J-1)^2 - z^2 J^2 \delta_\eta^2]}. \tag{37}$$

The planning count is finite for $J > 1$ and a positive denominator. At a fixed bin, the latter condition is $\delta_\eta < (J-1)/(zJ)$. A simultaneous fit of $\eta$ and the spin correlations uses their full joint covariance in place of the independent-error expression.

## 7 Numerical results and robustness

### 7.1 Steering after full angular averaging

We evaluate the averaged states using the BESIII central values [10]

$$\alpha = 0.4748, \quad \Delta\Phi = 0.7521\ \mathrm{rad}, \quad \alpha_\Lambda = 0.7519, \quad \alpha_{\bar{\Lambda}} = -0.7559, \tag{38}$$

giving $\eta = -0.56836121$. At $x = 0$, $F_3 = 1.20452$ and $G_3 = 1.12560$. The known-phase pointwise boundary is $|x| < 0.61974$, the all-phase guarantee is $|x| < 0.4748$, and the fixed-triad boundary is $|x| < 0.40952$.

Projective steering survives throughout the uniform central-bin range: its exact boundary $u_\mathrm{P} = 0.49723$ lies above the largest accessible moment $u = 1/3$. The CHSH and CJWR violation boundaries occur at $L = 0.61734$ and $L = 0.82238$, respectively. At the full-range endpoint $L = 1$,

$$F_3 = 0.94486247, \quad W_\mathrm{P} = 1.07013541, \quad R_\mathrm{crit} = 0.93446118. \tag{39}$$

The interval $0.82238 \leq L \leq 1$ thus contains angularly averaged states whose projective steering is missed by the three-setting witness.

### 7.2 Optimal angular selection and sample reduction

Minimizing Eq. (36) at nominal $z = 5$ gives

$$L_{\mathrm{opt},F} = 0.33684, \quad N_{\mathrm{full},F}^\mathrm{opt} = 8.6149\times 10^4, \quad L_{\mathrm{opt,P}} = 0.45592, \quad N_{\mathrm{full,P}}^\mathrm{opt} = 2.8047\times 10^4. \tag{40}$$

At the CJWR optimum, $F_3 = 1.16121$ and $f_\mathrm{acc} = 0.29604$, giving $25\,504$ selected pairs after upward rounding. At the projective-witness optimum, $W_\mathrm{P} = 1.27548$ and $f_\mathrm{acc} = 0.40657$, giving $11\,403$ selected pairs. The corresponding full-range targets are $86\,150$ and $28\,047$ pairs. With the same moment estimator and statistical model, the optimized projective witness reduces the required full-range sample by 67.4%, a factor of 3.07.

The projective witness combines a larger mean margin with a wider accepted angular interval. The variance coefficient $A_J$ in Eq. (35) completes the sample-cost calculation. Figure 3 and Table 1 compare the choices across bin widths. Using all events retains steering but requires about $1.78\times 10^5$ full-range pairs at the same statistical target; the optimized finite window lowers this requirement.

Table 1 Calculated full-range sample requirements for uniform central bins, the parameters in Eq. (38), known analyzing powers, and nominal $z = 5$. Counts are rounded upward. A dash marks bins without CJWR violation.

| $L$ | $F_3$ | $W_P$ | $f_{acc}$ | Full pairs CJWR | Full pairs projective |
|---|---|---|---|---|---|
| 0.20 | 1.18882 | 1.33682 | 0.17376 | 106,723 | 43,415 |
| 0.35 | 1.15792 | 1.30578 | 0.30803 | 86,303 | 29,887 |
| 0.50 | 1.11403 | 1.26117 | 0.44876 | 114,047 | 28,336 |
| 0.65 | 1.06194 | 1.20690 | 0.59871 | 290,935 | 34,093 |
| 0.75 | 1.02566 | 1.16768 | 0.70516 | 1,443,524 | 44,231 |
| 1.00 | 0.94486 | 1.07014 | 1.00000 | — | 177,768 |

The broad minima permit rounded angular selections with small sample penalties. Choosing $L = 0.35$ increases the CJWR full-range requirement by 0.18% relative to its optimum; $L = 0.20$ and $L = 0.50$ cost 23.9% and 32.4% more. For the projective witness, $L = 0.50$ is 1.03% above the optimum. These values quantify the sensitivity of the event requirement to rounding $L_{opt}$.

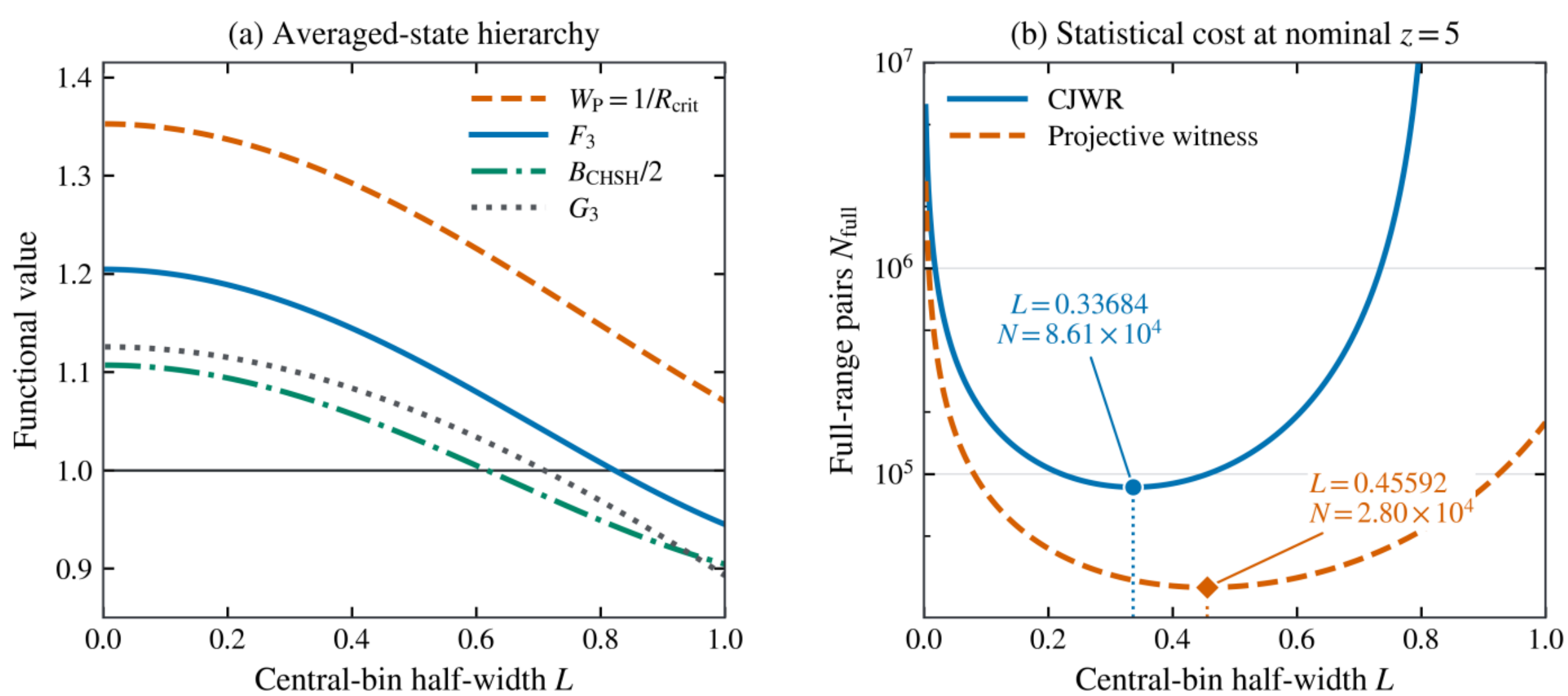

Fig. 3 Averaged-state witnesses and statistical cost. Panel (a) compares the projective functional, CJWR, fixed-triad witness, and one half of the maximal CHSH value, each with threshold one. Panel (b) gives the full-range pair requirement for the fixed linear witnesses at nominal $z = 5$. Dotted lines mark the separately optimized widths. The projective criterion detects steering through $L = 1$; the CJWR cost curve ends at its violation boundary.

Figure 4 scans the production parameter $\alpha$ with fixed analyzing powers. The black trajectories give the optimal central widths. The surfaces show how weaker production correlations raise the event requirement and how the projective criterion changes the preferred selection. The $\Lambda\overline{\Lambda}$ parameters in Eq. (38) define the reference point within this scan.

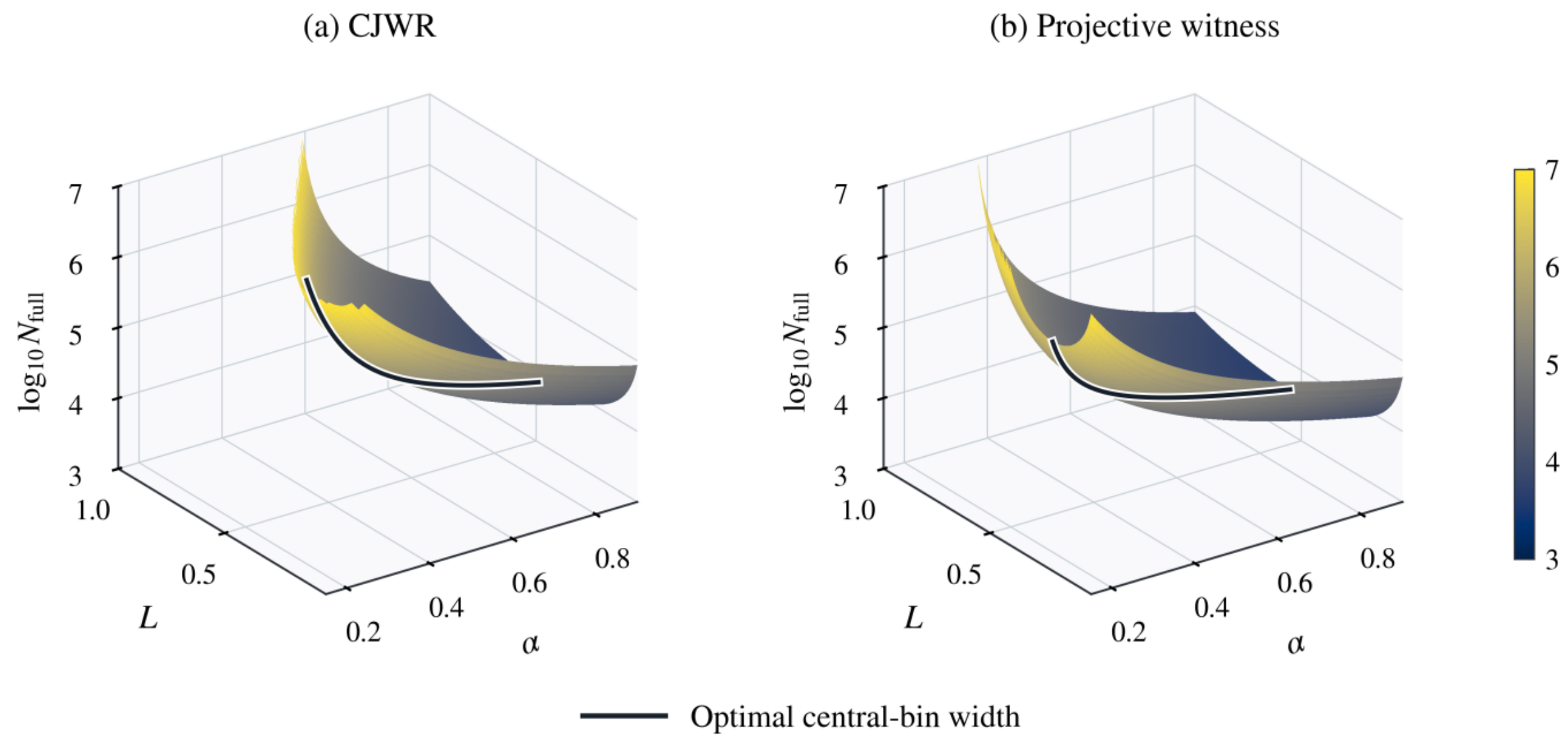

Fig. 4 Surfaces of $\log_{10} N_{\text{full}}(\alpha, L)$ for (a) CJWR and (b) the projective witness, with $\eta = -0.56836121$ and nominal $z = 5$. Both panels use the same height and color scales. Black curves give the minimizing bin at each $\alpha$. Surfaces are shown in the violation region for counts up to $10^7$ . Both horizontal axes are dimensionless; the vertical coordinate is the logarithm of the pair count.

### 7.3 Acceptance shape and parameter variations

We vary the even production-angle acceptance through the analytic shape family

$$\epsilon_\kappa(x) \propto 1 + \kappa x^2, \qquad -\tfrac{1}{2} \le \kappa \le \tfrac{1}{2}. \tag{41}$$

The overall normalization can be chosen so that $\epsilon_\kappa \le 1$; it cancels from the normalized moment and the selected fraction. With uniform daughter-angle acceptance held fixed, this variation isolates the effect of the production-angle response. The required quantities are

$$u_\kappa(L) = \frac{L^3/3 + \kappa L^5/5}{L + \kappa L^3/3}, \qquad v_\kappa = 0, \tag{42}$$

$$f_{\text{acc},\kappa}(L) = \frac{L + (\alpha+\kappa)L^3/3 + \alpha\kappa L^5/5}{1 + (\alpha+\kappa)/3 + \alpha\kappa/5}. \tag{43}$$

Equation (43) expresses the selected fraction relative to the reconstructed full-range sample with the same acceptance shape. We insert the corresponding $u$ and $f_{\text{acc}}$ into the steering and sample-size formulas to reoptimize each witness.

For $\kappa = -0.5$ and $+0.5$, the projective optimum moves to $L = 0.45367$ and $0.45809$. These shifts are below 0.5% relative to the uniform value. The corresponding full-range counts are approximately $2.33 \times 10^4$ and $3.26 \times 10^4$ , because the central fraction changes. The reduction relative to the separately optimized CJWR test remains about 67% in both cases.

External analyzing-power uncertainty has a larger effect on the event count. At $\delta_\eta = 1\%$, the optimized counts become approximately $9.87 \times 10^4$ for CJWR and $2.96 \times 10^4$ for the projective witness, increases of 14.5% and 5.6%. At 2%, they become $1.63 \times 10^5$ and $3.54 \times 10^4$. These scenarios specify external precision targets, with the central parameters held fixed. Table 2 and Fig. 5 give the reoptimized selections.

Table 2 Optimized full-range sample requirements under separate variations of external analyzing-power uncertainty or even production-angle acceptance. Unspecified variations are set to zero. Counts refer to reconstructed pairs for the stated acceptance and are rounded upward.

| Variation | $L_{\mathrm{opt},F}$ | Full pairs CJWR | $L_{\mathrm{opt,P}}$ | Full pairs projective |
|---|---|---|---|---|
| Baseline | 0.33684 | 86,150 | 0.45592 | 28,047 |
| $\delta_\eta = 0.5\%$ | 0.33270 | 89,021 | 0.45367 | 28,427 |
| $\delta_\eta = 1\%$ | 0.32015 | 98,666 | 0.44690 | 29,622 |
| $\delta_\eta = 2\%$ | 0.26682 | 162,874 | 0.41909 | 35,381 |
| $\kappa = -0.5$ | 0.33612 | 71,026 | 0.45367 | 23,345 |
| $\kappa = +0.5$ | 0.33755 | 100,930 | 0.45809 | 32,553 |

Varying $\alpha$ by $\sqrt{0.0022^2 + 0.0031^2} = 0.00380$, with the BESIII statistical and systematic errors combined in quadrature [10], gives projective optima $L = 0.45418$ – $0.45763$ and counts $2.73 \times 10^4$ – $2.88 \times 10^4$. Each shift defines a new benchmark state and its corresponding moment covariance. The symmetric average is independent of the phase because $\upsilon = 0$.

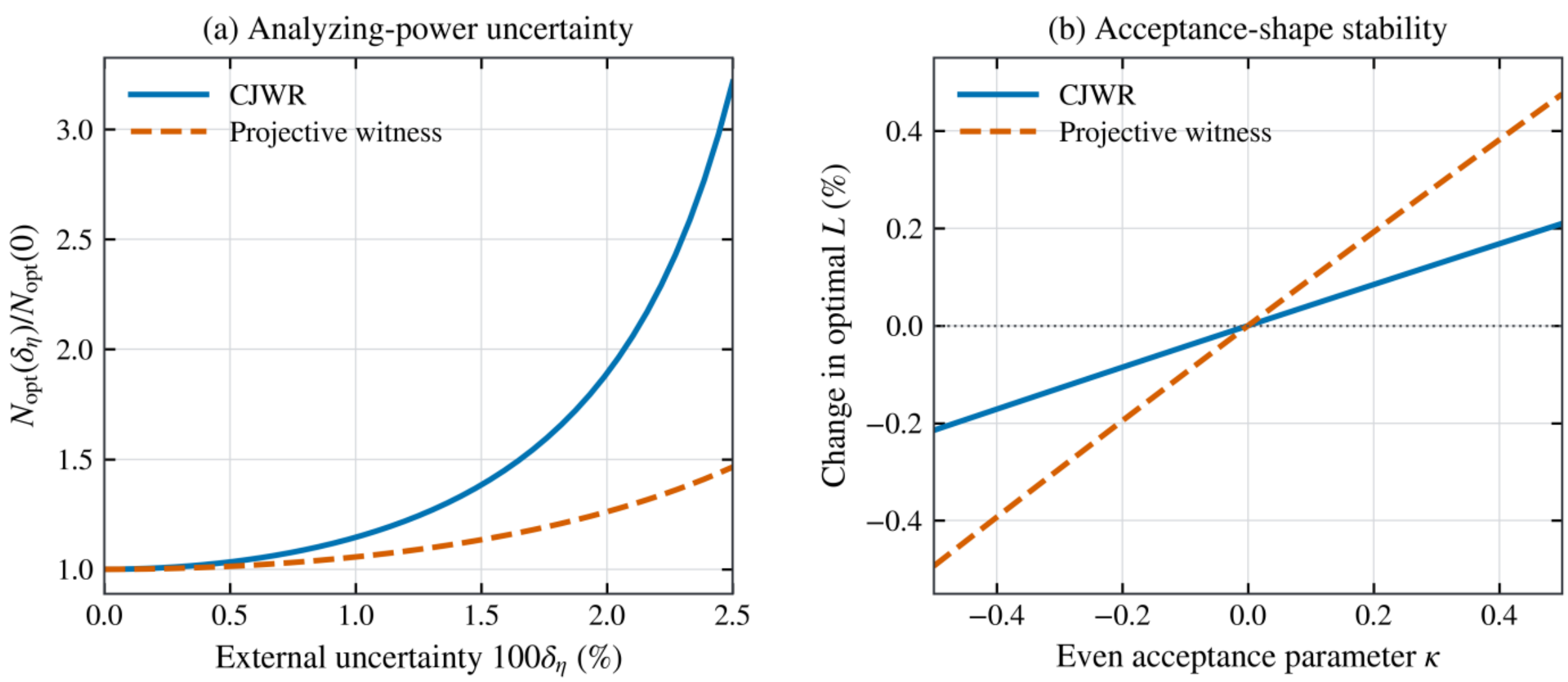

Fig. 5 Robustness of the optimized selection. Panel (a) shows the ratio of the reoptimized full-range pair count to its value at zero external analyzing-power uncertainty. Panel (b) shows the percentage change in the optimal half-width under the even acceptance family $\epsilon_\kappa \propto 1 + \kappa x^2$. The two criteria are reoptimized separately. These calculations retain uniform daughter-angle acceptance and the fixed production inputs of Eq. (38).

### 7.4 Numerical verification

Direct density-matrix checks over $6000$ random triples spanning both signs of $\alpha$ reproduce the pointwise spectrum and witness identities to floating-point precision. An additional $5000$ symmetric averaged states reproduce the Bell spectrum, CHSH threshold, and concurrence. Direct quadrature with $L = 0.67$ and the asymmetric acceptance $\epsilon(x) = 1 + 0.35x$ checks the nonzero-$\upsilon$ identity.

The full-range projective result and its critical $\alpha$ agree between $80 \times 160$ and $160 \times 320$ sphere grids. The inverse-tensor and norm integrals, as well as the elliptic-integral reduction, agree at the quoted precision. Numerical differentiation checks all nine entries of the projective gradient, and direct daughter-angle product quadrature reproduces its variance coefficient. A synthetic $400\,000$-pair sample also reproduces the CJWR linear variance coefficient, $26.5109$ compared with the analytic $26.5124$. Appendix D specifies these calculations and the parameter-grid convergence checks.

## 8 Connection to weak-decay measurements

Daughter directions resolved along the eventwise production axes reconstruct the averaged tensor in Eq. (11). The linear witnesses associated with $F_3$ and $W_{\mathrm{P}}$ use this same tensor, linking the steering boundaries directly to the event-level statistics.

Weak decays probe the parent spin through their analyzing powers. The daughter-direction measurement is described by the qubit positive-operator-valued measure [37,38]

$$E(d\Omega) = \frac{d\Omega}{4\pi}(I + a_Y \mathbf{n} \cdot \boldsymbol{\sigma}). \tag{44}$$

The measured decay directions determine the spin density matrix, from which we infer its steering capability under projective measurements. Classical binnings of these outcomes are jointly measurable postprocessings of a single measurement [39]. An operational steering protocol additionally requires independent choices of incompatible spin measurements. Reference [40] discusses this distinction for hyperon systems.

For data analysis, the angular bin and tangent-witness weights are determined from a production model or an independent design sample. A nonuniform joint efficiency $\epsilon(x, \mathbf{n}, \mathbf{m})$ enters an efficiency-normalized decay likelihood or a weighted estimator. Its covariance, including the effects of backgrounds and angular migration, supplies the uncertainty for the sample-size optimization. The variations in Sect. 7 quantify the separate effects of production-angle acceptance and analyzing-power precision.

## 9 Conclusions

We have determined the exact projective-steering boundary of symmetrically averaged hyperon states. In the production-frame spin convention, even angular acceptance cancels the local polarization and phase-dependent correlations, leaving a Bell-diagonal family. Applying its projective-steering criterion identifies a region beyond the three-setting test. For uniform full-range acceptance, the positive-$\alpha$ threshold is $0.36065$, compared with $1/\sqrt{3}$ for CJWR and $1/\sqrt{2}$ for CHSH.

At the $\Lambda\bar{\Lambda}$ benchmark, the full-range ensemble retains projective steering that CJWR misses. Combining the projective functional with the complete daughter-angle moment covariance yields the optimal central half-width $L = 0.45592$ and a full-range requirement of $2.80 \times 10^4$ pairs, compared with $L = 0.33684$ and $8.61 \times 10^4$ for CJWR at the same statistical target. The separately optimized projective witness therefore lowers the pair requirement by 67.4%. The reduction persists when the production-angle acceptance and external analyzing-power uncertainty are varied.

The fixed linear witnesses connect the projective-steering boundary to event weights with explicit means and variances. Together with the accepted yield, they turn angular selection into a quantitative sample-size optimization for hyperon spin measurements.

## Acknowledgements

This work was supported by the Key Teaching Research Project of the 2025 University-Level Quality Engineering Program at Bengbu Medical University (Grant No. 2025jyxm08), the Anhui Provincial University Scientific Research Key Project (Grant No. 2025AHGXZK30635) and the Natural Science Foundation of Anhui Sanlian University (Grant Nos. KJZD2025021 and KJZD2025020).

## Data availability

The production and decay inputs are published in Ref. [10]. All figures and tables are theoretical calculations; the numerical sampling check uses synthetic decay directions. The equations, numerical settings, and reproduction procedure are given in this article, and the Online Resource 1 provides the tabulated figure data and calculation and plotting code.

## Appendix A: Pointwise X form and CJWR optimization

At fixed $x$, a local unitary maps Eq. (2) to an X state. Introduce

$$\mathrm{D}_X = [1 + \alpha(2x^2 - 1)]^2 - 4\beta^2 x^2 (1 - x^2), \quad p = \frac{\beta x s}{d}, \tag{A1}$$

$$t_{1,2} = \frac{1 + \alpha \pm \sqrt{\mathrm{D}_X}}{2d}, \quad t_3 = -\frac{\alpha(1 - x^2)}{d}. \tag{A2}$$

Then

$$\rho_X = \frac{1}{4}\begin{pmatrix} 1 + 2p + t_3 & 0 & 0 & t_1 - t_2 \\ 0 & 1 - t_3 & t_1 + t_2 & 0 \\ 0 & t_1 + t_2 & 1 - t_3 & 0 \\ t_1 - t_2 & 0 & 0 & 1 - 2p + t_3 \end{pmatrix}. \tag{A3}$$

These formulas agree with Ref. [18]. On the positive branch, $t_1 \geq t_2 \geq 0$ and $t_3 \leq 0$. The discriminant $\mathrm{D}_X$ is distinct from the steering critical radius $R_{\mathrm{crit}}$.

For a fixed orthonormal Bob triad, Alice chooses $\mathbf{a}_k$ along $C\mathbf{b}_k$, giving

$$\max_{\{\mathbf{a}_k\}} f_3 = \frac{1}{\sqrt{3}} \sum_k \| C\mathbf{b}_k \| \leq \sqrt{\sum_k \| C\mathbf{b}_k \|^2} = \| C \|_{\mathrm{F}}. \tag{A4}$$

To attain the bound, choose Bob's triad so that each diagonal element of the rotated matrix $C^{\mathrm{T}}C$ equals one third of its trace. Such a real basis exists: choose a first direction whose Rayleigh quotient equals the mean eigenvalue, then rotate the principal directions in its orthogonal plane by 45°. All three norms in Eq. (A4) are then equal. Alice's resulting directions need not be orthogonal.

Restricting Bob instead to principal axes yields $(|t_1| + |t_2| + |t_3|)/\sqrt{3}$. Equality with $F_3$ requires equal singular values, apart from the trivial zero tensor. This explains the distinction between the optimized norm and the restricted principal-axis sum.

## Appendix B: Two-moment identity and acceptance symmetry

Let $D_B = 1 + \alpha u$. Squaring the entries of Eq. (14) gives

$$D_B^2 F_3(\overline{\rho})^2 = (1+\alpha^2)(1-u)^2 + (\alpha+u)^2 + 2\gamma^2 v^2. \quad \text{(B1)}$$

The polynomial identity

$$(1+\alpha^2)(1-u)^2 + (\alpha+u)^2 - (1+\alpha u)^2 = 2(1-u)(\alpha^2-u) \quad \text{(B2)}$$

proves Eq. (15). Reflection changes the sign of $x\sqrt{1-x^2}$ and leaves all diagonal numerators even. Pairing reflected angles with equal acceptance therefore gives $v = 0$. Cauchy–Schwarz applied to $x$ and $\sqrt{1-x^2}$ under the normalized acceptance measure gives $v^2 \le u(1-u)$.

The same averaged tensor gives

$$G_3(\overline{\rho}) = \frac{1+2\alpha-\alpha u}{\sqrt{3}(1+\alpha u)}. \quad \text{(B3)}$$

This expression is independent of $v$ even for asymmetric acceptance. Its phase independence therefore has a different origin from that of the optimized norm in a symmetric bin.

## Appendix C: Projective functional and statistical weights

### C.1 Sphere integral and the local-hidden-state bound

For Bob's direction $\mathbf{n}$, choose Alice's projective direction along $C\mathbf{n}$. The quantum correlation integrated uniformly over Bob's sphere gives $W_\mathrm{P}(C)$. For a hidden state with Bloch vector $\mathbf{r}_\lambda$ and an arbitrary response $A_\lambda(\mathbf{n}) \in [-1,1]$, its contribution is bounded by

$$\frac{1}{2\pi}\int_{S^2} d\,\Omega\, A_\lambda(\mathbf{n})\mathbf{r}_\lambda \cdot \mathbf{n} \le \frac{1}{2\pi}\int_{S^2} d\,\Omega\, |\mathbf{r}_\lambda \cdot \mathbf{n}| = |\mathbf{r}_\lambda| \le 1. \quad \text{(C1)}$$

Convex mixing preserves this bound. Hence the sphere integral is a sufficient steering witness for a general tensor. For zero-marginal T states it coincides with the exact critical-radius criterion [24–26].

For invertible diagonal $T$, set $\mathbf{n} = T\mathbf{m}/\| T\mathbf{m} \|$. The solid-angle Jacobian is $d\Omega_n = |\det T| \, \| T\mathbf{m} \|^{-3}\, d\Omega_m$, while $\mathbf{n}^\mathrm{T} T^{-2}\mathbf{n} = \| T\mathbf{m} \|^{-2}$. Equation (20) therefore reduces to Eq. (21). The norm integral extends continuously to singular tensors.

For the positive branch, $c_1^2 \ge c_2^2$. Integrating the azimuth analytically gives

$$W_\mathrm{P}(T) = \frac{4}{\pi}\int_0^1 d\,z \sqrt{A(z)}\, E\left(\frac{B(z)}{A(z)}\right), \quad \text{(C2)}$$

where $A(z) = c_1^2(1-z^2) + c_3^2 z^2$, $B(z) = (c_1^2 - c_2^2)(1-z^2)$, and $E(m) = \int_0^{\pi/2} \sqrt{1 - m\sin^2\phi}\; d\phi$ is the complete elliptic integral with parameter $m$. This form is convenient near the rank-one endpoint.

To establish Eq. (25), let $\lambda_i$ be the squared singular values divided by $F_3^2$. The sphere mean of $\sqrt{\sum_i \lambda_i n_i^2}$ is concave on $\sum_i \lambda_i = 1$. Its minimum is the rank-one value $\langle |n_1| \rangle = 1/2$, which proves the lower bound. Jensen's inequality gives the upper bound because $\langle n_i^2 \rangle = 1/3$. Strict concavity gives the stated strictness for rank greater than one.

### C.2 Tangent statistics and norm bias

Differentiating the integral gives Eq. (31). At diagonal $T$, the off-diagonal derivatives vanish and

$$(K_\mathrm{P})_{ii} = \frac{1}{2\pi}\int_{S^2} d\,\Omega\, \frac{c_i n_i^2}{\sqrt{\sum_k c_k^2 n_k^2}}. \quad \text{(C3)}$$

For a fixed design matrix $K_J(C_0)$, each pair contributes

$$Z_J^{(r)} = \frac{9}{\eta}\, \mathbf{n}^{(r)\mathrm{T}} K_J(C_0) \mathbf{m}^{(r)}, \quad \hat{S}_J = \frac{1}{N}\sum_r Z_J^{(r)}. \quad \text{(C4)}$$

Convexity gives $J(C) \ge J(C_0) + K_J(C_0) : (C - C_0)$, and homogeneity cancels the constant term, proving Eq. (33). Contracting the complete covariance with the fixed matrix gives the exact finite-$N$ variance in Eq. (34).

The leading biases of the nonlinear estimates $J(\hat{C})$ follow from their Hessians. For a nonsingular true tensor,

$$\mathrm{Bias}(\hat{F}_3) \simeq \frac{36}{N\eta^2 F_3}, \quad \mathrm{Bias}(\widehat{W}_\mathrm{P}) \simeq \frac{18}{N\eta^2}\left\langle \frac{1}{\|C\mathbf{n}\|} \right\rangle_{S^2}. \quad \text{(C5)}$$

The identities behind these expressions are $\mathrm{Tr} H_F = 8/F_3$, $\mathrm{Tr} H_\mathrm{P} = 4\langle \| C\mathbf{n} \|^{-1} \rangle$, and $\mathbf{c}^\mathrm{T} H_J \mathbf{c} = 0$. At the respective benchmark selected-pair counts, the biases are approximately $0.00376$ and $0.00799$. The linear statistics in Eq. (C4) estimate their witness means without bias.

## Appendix D: Numerical reproduction

The reported numerical results use double-precision NumPy and SciPy. Online Resource 1 includes the Python implementation of the equations, figure data, and numerical validation outputs. The two-dimensional sphere quadrature uses Gauss–Legendre nodes in $z = \cos\vartheta$ and an equally spaced azimuth: the baseline grid has $80 \times 160$ nodes and the comparison grid $160 \times 320$. The implementation normalizes the sphere weights to unity; a solid-angle integral is $4\pi$ times the corresponding weighted mean. The elliptic representation uses 96 Gauss–Legendre nodes on $[0,1]$. Repeating the projective-boundary calculation with 192 nodes on a 60-point grid over $0.015 \le \alpha \le 0.985$ changes $u_\mathrm{P}$ by less than $1.4 \times 10^{-8}$.

The full-range critical $\alpha$ is obtained by bracketing the root of $W_\mathrm{P}[T(\alpha, 1/3)] - 1$. The inverse-tensor integral and the norm integral give $R_{\text{crit}} = 0.9344611816$ at the benchmark. Both sphere resolutions give $\alpha_{\text{P}}(1) = 0.3606459793$ at the displayed precision. The elliptic and sphere formulas agree within $3 \times 10^{-14}$ for the quoted benchmark value and gradient.

We evaluate the full-range planning count at 301 widths, bracket its lowest finite value by the neighboring widths, and apply bounded scalar minimization with absolute tolerance $10^{-12}$ in $L$. The endpoint $L = 1$ is included whenever the witness violates there. The minimization uses bins with a positive violation margin and a positive uncertainty denominator. We repeat this procedure for each acceptance and external-error scenario.

For the direct state check, we initialize NumPy's default random generator with seed 20260920 and sample 6000 independent triples uniformly in $\alpha \in [-1,1]$, $x \in [-1,1]$, and $\Delta\Phi \in [-\pi, \pi]$. We construct Eq. (2) with Pauli matrices and compare its spectrum and Frobenius norm with the analytic expressions. Continuing the generator with 5000 pairs $(\alpha, u)$ in $(0,1) \times [0,1)$, we compare the Bell spectrum, the two largest eigenvalues of $T^{\text{T}}T$, and twice the partial-transpose negativity with Eqs. (19), (22), and (23). The spectral and witness residuals remain below $10^{-14}$.

For the projective gradient, central differences of step $10^{-5}$ in each of the nine tensor entries at $\alpha = 0.4748$, $u = 0.18$ agree with the integral gradient within $5 \times 10^{-11}$. Euler's relation $K_{\text{P}}\!:\!T = W_{\text{P}}$ agrees within $2 \times 10^{-14}$. A product of two $14 \times 28$ daughter-sphere quadratures gives the projective linear-score variance coefficient $35.06288693$, equal to the contraction of Eq. (30) at the quoted precision.

The synthetic CJWR check uses a separate generator with seed 20260908. After consuming 6000 three-component uniform draws, we generate independent three-component standard normal vectors in batches of 200000 and normalize them to obtain proposed daughter directions. We use the tensor at $L = 0.3368416818$ and accept each proposal with probability

$$P_{\text{keep}} = \frac{1+\eta\, \mathbf{n}^{\text{T}} T \mathbf{m}}{1+|\eta|\, \| T \|_2}, \tag{D1}$$

where $\| T \|_2$ is the spectral norm. We retain the first 400000 accepted pairs. The unbiased sample variance of $9\mathbf{n}^{\text{T}}(T/F_3)\mathbf{m}/\eta$ is $26.51087594$, compared with the analytic coefficient $26.51239258$. This agreement checks the event-level second moment used in the sample-size calculation.

## Statements and Declarations

**Competing interests.** The authors declare no competing interests.

## References

[1] A. Einstein, B. Podolsky, and N. Rosen, Can quantum-mechanical description of physical reality be considered complete? Phys. Rev. **47**, 777–780 (1935). https://doi.org/10.1103/PhysRev.47.777.

[2] E. Schrödinger, Discussion of probability relations between separated systems, Proc. Cambridge Philos. Soc. **31**, 555–563 (1935). https://doi.org/10.1017/S0305004100013554.

[3] H. M. Wiseman, S. J. Jones, and A. C. Doherty, Steering, entanglement, nonlocality, and the Einstein–Podolsky–Rosen paradox, Phys. Rev. Lett. **98**, 140402 (2007). https://doi.org/10.1103/PhysRevLett.98.140402.

[4] S. J. Jones, H. M. Wiseman, and A. C. Doherty, Entanglement, Einstein–Podolsky–Rosen correlations, Bell nonlocality, and steering, Phys. Rev. A **76**, 052116 (2007). https://doi.org/10.1103/PhysRevA.76.052116.

[5] R. Uola, A. C. S. Costa, H. Chau Nguyen, and O. Gühne, Quantum steering, Rev. Mod. Phys. **92**, 015001 (2020). https://doi.org/10.1103/RevModPhys.92.015001.

[6] A. J. Barr, M. Fabbrichesi, R. Floreanini, E. Gabrielli, and L. Marzola, Quantum entanglement and Bell inequality violation at colliders, Prog. Part. Nucl. Phys. **139**, 104134 (2024). https://doi.org/10.1016/j.ppnp.2024.104134.

[7] G. Fäldt and A. Kupść, Hadronic structure functions in the $e^+e^- \to \bar{\Lambda}\Lambda$ reaction, Phys. Lett. B **772**, 16–20 (2017). https://doi.org/10.1016/j.physletb.2017.06.011.

[8] E. Perotti, G. Fäldt, A. Kupść, S. Leupold, and J. J. Song, Polarization observables in $e^+e^-$ annihilation to a baryon–antibaryon pair, Phys. Rev. D **99**, 056008 (2019). https://doi.org/10.1103/PhysRevD.99.056008.

[9] M. Ablikim et al. (BESIII Collaboration), Polarization and entanglement in baryon–antibaryon pair production in electron–positron annihilation, Nat. Phys. **15**, 631–634 (2019). https://doi.org/10.1038/s41567-019-0494-8.

[10] M. Ablikim et al. (BESIII Collaboration), Precise measurements of decay parameters and CP asymmetry with entangled $\Lambda$–$\bar{\Lambda}$ pairs, Phys. Rev. Lett. **129**, 131801 (2022). https://doi.org/10.1103/PhysRevLett.129.131801.

[11] C. Qian, J.-L. Li, A. S. Khan, and C.-F. Qiao, Nonlocal correlation of spin in high energy physics, Phys. Rev. D **101**, 116004 (2020). https://doi.org/10.1103/PhysRevD.101.116004.

[12] S. Wu, C. Qian, Q. Wang, and X.-R. Zhou, Bell nonlocality and entanglement in $e^+e^- \to Y\bar{Y}$ at BESIII, Phys. Rev. D **110**, 054012 (2024). https://doi.org/10.1103/PhysRevD.110.054012.

[13] M. Fabbrichesi, R. Floreanini, E. Gabrielli, and L. Marzola, Bell inequality is violated in charmonium decays, Phys. Rev. D **110**, 053008 (2024). https://doi.org/10.1103/PhysRevD.110.053008.

[14] Y. Afik and J. R. Muñoz de Nova, Entanglement and quantum tomography with top quarks at the LHC, Eur. Phys. J. Plus **136**, 907 (2021). https://doi.org/10.1140/epjp/s13360-021-01902-1.

[15] Y. Afik and J. R. Muñoz de Nova, Quantum information with top quarks in QCD, Quantum **6**, 820 (2022). https://doi.org/10.22331/q-2022-09-29-820.

[16] E. G. Cavalcanti, S. J. Jones, H. M. Wiseman, and M. D. Reid, Experimental criteria for steering and the Einstein–Podolsky–Rosen paradox, Phys. Rev. A **80**, 032112 (2009). https://doi.org/10.1103/PhysRevA.80.032112.

[17] A. C. S. Costa and R. M. Angelo, Quantification of Einstein–Podolsky–Rosen steering for two-qubit states, Phys. Rev. A **93**, 020103(R) (2016). https://doi.org/10.1103/PhysRevA.93.020103.

[18] S. Wu, C. Qian, Q. Wang, and Y.-G. Yang, Quantum steering and discord in hyperon–antihyperon system in electron–positron annihilation, Phys. Rev. D **113**, 056030 (2026). https://doi.org/10.1103/x43x-3wgy.

[19] K. Cheng, T. Han, and M. Low, Optimizing fictitious states for Bell inequality violation in bipartite qubit systems with applications to the $t\bar{t}$ system, Phys. Rev. D **109**, 116005 (2024). https://doi.org/10.1103/PhysRevD.109.116005.

[20] D. Cavalcanti and P. Skrzypczyk, Quantum steering: A review with focus on semidefinite programming, Rep. Prog. Phys. **80**, 024001 (2017). https://doi.org/10.1088/1361-6633/80/2/024001.

[21] S. Jevtic, M. Pusey, D. Jennings, and T. Rudolph, Quantum steering ellipsoids, Phys. Rev. Lett. **113**, 020402 (2014). https://doi.org/10.1103/PhysRevLett.113.020402.

[22] H. Chau Nguyen and T. Vu, Nonseparability and steerability of two-qubit states from the geometry of steering outcomes, Phys. Rev. A **94**, 012114 (2016). https://doi.org/10.1103/PhysRevA.94.012114.

[23] S. Jevtic, M. J. W. Hall, M. R. Anderson, M. Zwierz, and H. M. Wiseman, Einstein-Podolsky-Rosen steering and the steering ellipsoid, J. Opt. Soc. Am. B **32**, A40–A49 (2015). https://doi.org/10.1364/JOSAB.32.000A40.

[24] H. Chau Nguyen and T. Vu, Necessary and sufficient condition for steerability of two-qubit states by the geometry of steering outcomes, EPL **115**, 10003 (2016). https://doi.org/10.1209/0295-5075/115/10003.

[25] H. Chau Nguyen, Huy-Viet Nguyen, and O. Gühne, Geometry of Einstein-Podolsky-Rosen correlations, Phys. Rev. Lett. **122**, 240401 (2019). https://doi.org/10.1103/PhysRevLett.122.240401.

[26] Y. Afik and J. R. Muñoz de Nova, Quantum discord and steering in top quarks at the LHC, Phys. Rev. Lett. **130**, 221801 (2023). https://doi.org/10.1103/PhysRevLett.130.221801.

[27] B. Jungnitsch, S. Niekamp, M. Kleinmann, O. Gühne, H. Lu, W.-B. Gao, Y.-A. Chen, Z.-B. Chen, and J.-W. Pan, Increasing the statistical significance of entanglement detection in experiments, Phys. Rev. Lett. **104**, 210401 (2010). https://doi.org/10.1103/PhysRevLett.104.210401.

[28] C. Schwemmer, L. Knips, D. Richart, H. Weinfurter, T. Moroder, M. Kleinmann, and O. Gühne, Systematic errors in current quantum state tomography tools, Phys. Rev. Lett. **114**, 080403 (2015). https://doi.org/10.1103/PhysRevLett.114.080403.

[29] J. S. Bell, On the Einstein Podolsky Rosen paradox, Physics Physique Fizika **1**, 195–200 (1964). https://doi.org/10.1103/PhysicsPhysiqueFizika.1.195.

[30] J. F. Clauser, M. A. Horne, A. Shimony, and R. A. Holt, Proposed experiment to test local hidden-variable theories, Phys. Rev. Lett. **23**, 880–884 (1969). https://doi.org/10.1103/PhysRevLett.23.880.

[31] R. Horodecki, P. Horodecki, and M. Horodecki, Violating Bell inequality by mixed spin-1/2 states: necessary and sufficient condition, Phys. Lett. A **200**, 340–344 (1995). https://doi.org/10.1016/0375-9601(95)00214-N.

[32] W. K. Wootters, Entanglement of formation of an arbitrary state of two qubits, Phys. Rev. Lett. **80**, 2245–2248 (1998). https://doi.org/10.1103/PhysRevLett.80.2245.

[33] Q. Quan, H. Zhu, S.-Y. Liu, S.-M. Fei, H. Fan, and W.-L. Yang, Steering Bell-diagonal states, Sci. Rep. **6**, 22025 (2016). https://doi.org/10.1038/srep22025.

[34] A. Bernal, Quantum tomography of helicity states for general scattering processes, Phys. Rev. D **109**, 116007 (2024). https://doi.org/10.1103/PhysRevD.109.116007.

[35] R. Ashby-Pickering, A. J. Barr, and A. Wierzchucka, Quantum state tomography, entanglement detection and Bell violation prospects in weak decays of massive particles, J. High Energy Phys. 05 (2023) 020. https://doi.org/10.1007/JHEP05(2023)020.

[36] R. Ashby-Pickering, A. J. Barr, and A. Wierzchucka, Erratum to: Quantum state tomography, entanglement detection and Bell violation prospects in weak decays of massive particles, J. High Energy Phys. 03 (2026) 166. https://doi.org/10.1007/JHEP03(2026)166.

[37] B. C. Hiesmayr, Limits of quantum information in weak interaction processes of hyperons, Sci. Rep. **5**, 11591 (2015). https://doi.org/10.1038/srep11591.

[38] S. Wu, C. Qian, Y.-G. Yang, and Q. Wang, Generalized quantum measurement in spin-correlated hyperon–antihyperon decays, Chin. Phys. Lett. **41**, 110301 (2024). https://doi.org/10.1088/0256-307X/41/11/110301.

[39] R. Uola, C. Budroni, O. Gühne, and J.-P. Pellonpää, One-to-one mapping between steering and joint measurability problems, Phys. Rev. Lett. **115**, 230402 (2015). https://doi.org/10.1103/PhysRevLett.115.230402.

[40] S. Haddadi, Comment on "Controlling the dynamical evolution of quantum coherence and quantum correlations in $e^+e^- \to \Lambda\bar{\Lambda}$ processes at BESIII", Phys. Rev. D **114**, 038901 (2026). https://doi.org/10.1103/hsvc-9cn1.